\pdfoutput =1
\documentclass{JINST}
\let\ifpdf\relax
\usepackage{multirow}
\usepackage{setspace}

\title{Evaluation of  commercial ADC radiation tolerance for accelerator experiments }

\author{Kai Chen,~Hucheng Chen,~James Kierstead, ~Helio Takai\thanks{Corresponding
author.},   ~Sergio Rescia, ~Xueye Hu, 
~Hao Xu, ~Joseph Mead, ~Francesco Lanni and Marena Minelli\\  
\llap {}Brookhaven National Laboratory,\\
  PO Box 5000, Upton, NY 11973, USA\\
  E-mail: \email{takai@bnl.gov}}

\abstract{Electronic components used in high energy physics experiments are subjected to a radiation background composed of high energy hadrons, mesons and photons. These particles can induce permanent and transient effects that affect the normal device operation. Ionizing dose and displacement damage can cause chronic damage which
disable the device permanently. Transient effects or single event effects are in general recoverable with time intervals that depend on the nature of the failure. The magnitude of these effects is technology dependent with feature size being one of the key parameters.  
Analog to digital converters
 are components that are frequently used in detector front end electronics, generally placed as close
as possible to the sensing elements to maximize signal fidelity. We report on radiation effects tests conducted on 17 commercially available analog to digital converters and extensive single event effect measurements on specific twelve and fourteen bit ADCs that presented
high tolerance to ionizing dose. Mitigation strategies for
single event effects (SEE) are discussed for their use in the large hadron collider environment. }


\begin{document}

\section{Introduction}

Electronic components used in high energy physics accelerator facilities are subjected to a unique radiation background
that is generated by the accelerator itself.  Generated by particles from collisions or unwanted beam interactions 
with the accelerator instrumentation, this 
man made radiation is composed of high energy hadrons, mesons and photons. 
The nature of particle production makes this background free from heavy ions and
with similar characteristics
to terrestrial cosmic ray generated radiation but at a much higher flux.  
The exact particle composition and rates 
depend on the amount of material in the vicinity of
the electronics and most importantly between
the electronics and the interaction point. 
At collider facilities the majority of particles impinging on the electronics near the interaction point are 
mesons ($\pi$,$K$, etc..). Further out, neutrons and gamma rays dominate the background. 

An important electronic component for many applications 
is an analog to digital converter. Digitizing analog signals
as close as possible to the detector and transporting them via optical fibers  
guarantees signal fidelity,  especially if the alternative is to use long cable runs. 
Typically, a collider experiment requires the digitization of many thousands of channels at
frequencies that vary from 10 to 100~MHz  with a dynamic range of 10 to 16 bits. 
A design that uses commercial off the shelf (COTS) components can lead to 
faster electronics development, flexibility and may represent cost saving for future 
detector electronics.
As the feature size  of integrated circuits 
decreases the observed trend for some time has been an increase in ionizing
radiation resistance \cite{rad1}. 
On the other hand the susceptibility to single event effects increases as less charge is required to 
switch a transistor on or off.  Hence, an evaluation of the radiation tolerance of commercial off
the shelf ADCs is merited. 

In this paper we report on results of ionizing dose tests of seventeen commercial ADCs and on results of
an extensive single event effect studies performed on two components with different dynamic ranges.  
Based on the
results obtained we discuss possible strategies for their use in the large hadron collider radiation environment.



\section{A unique radiation environment}

The accelerator radiation environment
is anthropogenic, with damaging radiation generally existing only
when the accelerator is in operation. Every accelerator
has yearly scheduled operations and unscheduled down time 
leading to an episodical radiation exposure of components. 
When the accelerator is operating
interactions happen following a
well defined periodical timing structure that includes
intervals that are free of particle collisions. The component
exposure is subjected to the accelerator duty cycle 
making changes in performance due to annealing
important to be studied. 
In addition, during long scheduled downtimes 
the electronics can be accessed for repairs, if required.

\begin{figure}
\centering
\includegraphics[width=0.8\columnwidth]{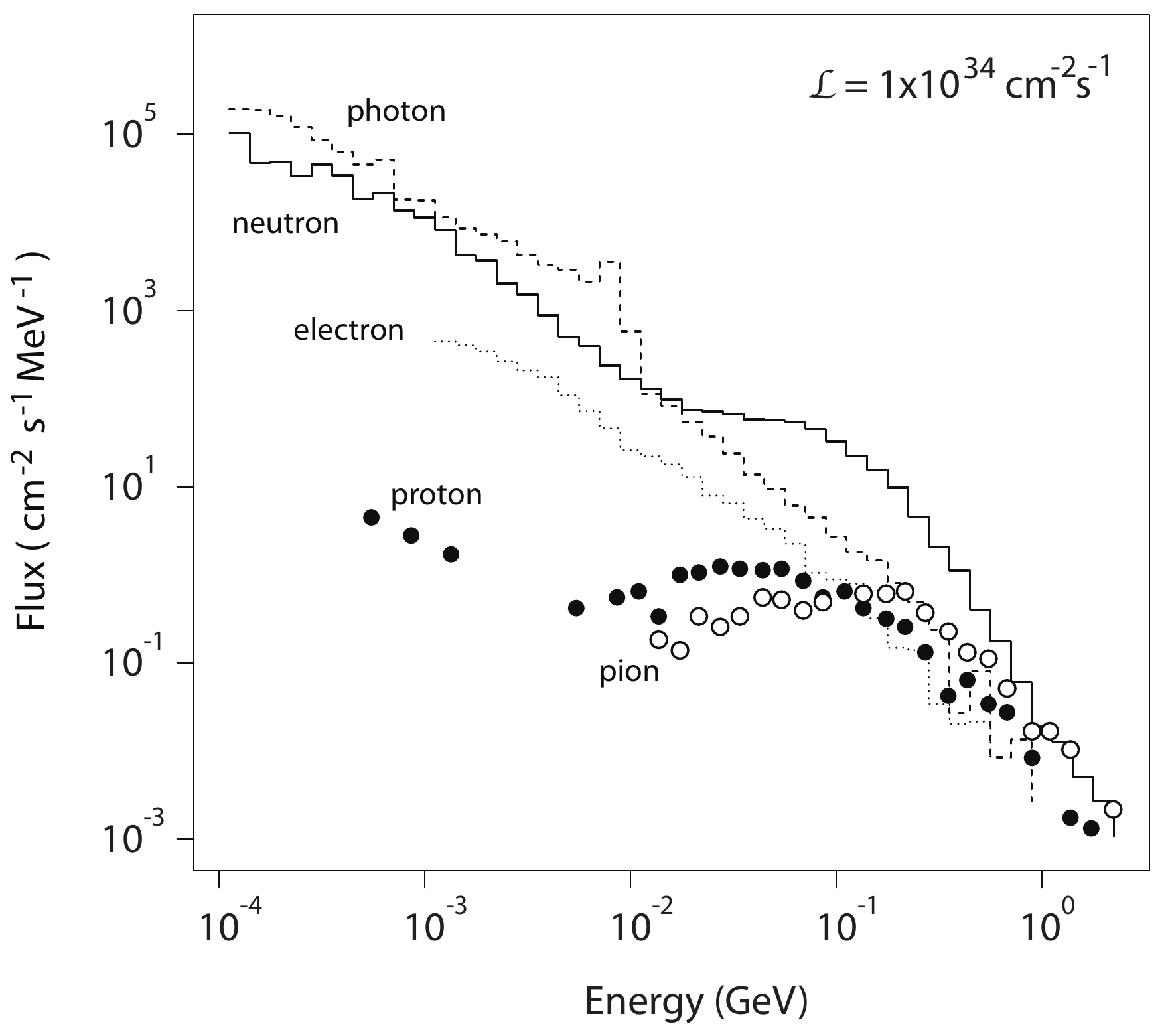}
\caption{ Energy spectra of particles in the location of the ATLAS liquid argon calorimeter electronics. The environment is
composed by hadrons, mesons and photons.}
\label{fig:radfield}
\end{figure}

Test and qualification of components to be used in the accelerator
radiation environment should take the radiation environment characteristics
into account. 
The background composition
is a function
of where the component is placed with 
respect to the collision point. Material type and thickness before components
defines how many interactions 
high energy particles
undergo before they reach them determining the
population of particles and their energy spectra. Components
placed near the collision point are subjected mostly to 
mesons ($\pi$,$K$, etc..) that deposit dose by the
energy loss process as well as induce
single event effects through nuclear fragmentation reactions such as $^{28}Si(\pi , X)$. 
In contrast, components in
the external layers of a detector are struck
mostly by neutrons that are produced at the later stages of
the particle shower. 
Fig \ref{fig:radfield}, illustrates the expected particle composition behind the ATLAS
detector calorimeters \cite{shupe}. The neutron energy spectrum shows a close resemblance in shape
to the atmospheric neutron spectrum at sea level due to the similar amount of material 
for particle interactions \cite{ibmneutron}.  The background also contains
electrons, photons, protons, and
pions. The overall flux is $\sim 10^6$ times higher than the terrestrial radiation. 

Our work concentrates on the
qualification of components in the environment 
behind calorimeters, 
that is fairly well shielded from ionizing radiation. 
At these locations components are subjected to
low total ionizing dose. 
The dose rate
is low and effects such as extreme low dose radiation sensitivity
(ELDRS) become relevant. Single event effects are
induced by neutrons with energy greater than 10~MeV
causing
nuclear reactions such as $^{28}Si(n,\alpha)$ and $^{28}Si(n,X)$ in the
semiconductor media. A second possible source of single event effects
are due to thermal neutrons that originate from the thermalization of neutrons 
in the heavily shielded experimental hall. The main nuclear
interaction is the neutron capture reaction with the $^{10}B$ isotope found
in the p-type silicon. This reaction produced two fragments, $\alpha$ and  $^7 Li$ that
are able to induce SEE. Permanent damage can also be
caused by the non ionizing energy loss that induces displacement damage. However,
this is not relevant for CMOS devices until very large fluences. 


\section{Test Setup}

\begin{figure}
\centering
\includegraphics[width=0.9\columnwidth]{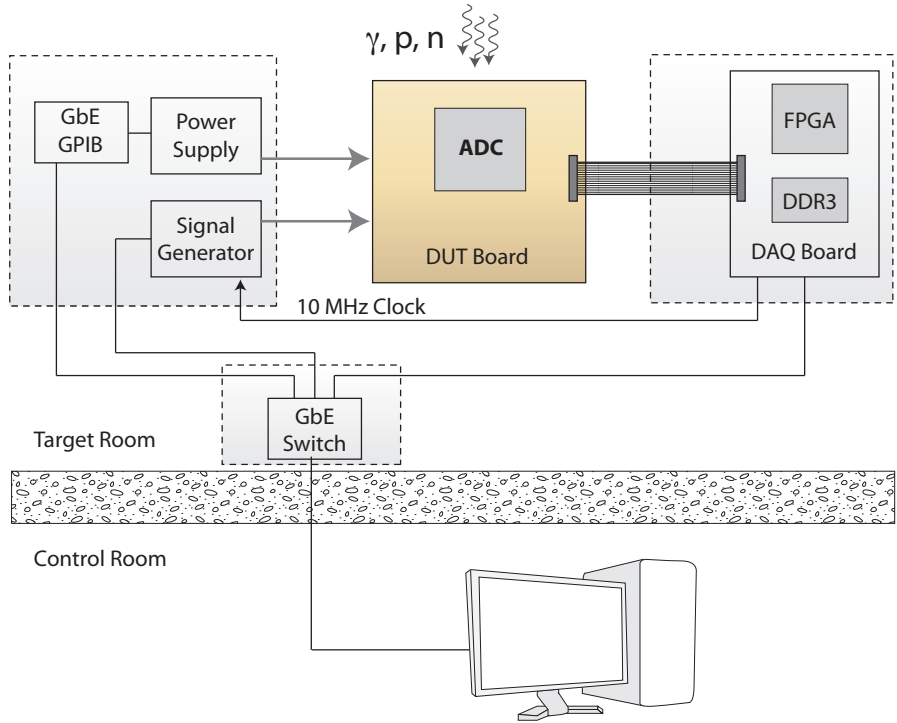}
\caption{Experimental setup for the dynamic test of ADCs for radiation effects. The device under test (DUT) is mounted on the
DUT board and connected to the data acquisition (DAQ) board via a short high speed cable. A sinusoidal waveform of 40~kHz 
is fed to the ADC and its output compared sample by sample with the expected output. Upon the detection of an inconsistency
samples are recorded by the main data acquisition computer for offline analysis. Shaded squares on the components in the Target Room
indicate shielding material used to prevent radiation to reach them. 
}
\label{fig:setup}
\end{figure}

To perform ADC radiation tests we designed and implemented 
a setup that allows for the dynamical
test of ADCs. The setup
allows us to examine the  ADC  output sample by sample to detect
radiation induced data modification. 
A block diagram
with the relevant components is shown in Fig. \ref{fig:setup}. The device under test 
(DUT) that is exposed to various radiation sources
is mounted on
the DUT board.  
Data is acquired by
a data acquisition board (DAQ) that is shielded
from charged particles and neutrons 
by a combination of lead and polyethylene blocks. A short high
speed extension cable connects the DAQ and DUT boards for
fast data transfer. 
The power supply and
signal generator for the device under test are also located in the target room
behind shielding. The signal generator is 
synchronized to the DAQ board.  
All devices are controlled by a main computer located in the control room via
 gigabit ethernet connections.

A sine wave of
frequency $f_{0}= 40~kHz$ and $V_{pp}=2~V$  is fed to the ADC and digitized at 
40$Ms \cdot s^{-1}$. To detect the presence of an erroneous output signal, e.g. caused by
a single event effect,  the digitized signal is compared sample by sample to the expected 
values obtained from a lookup table.  Prior to irradiation the lookup table is generated
by an iterative routine which averages digital data to obtain a reference waveform. Later
the phase and amplitude are adjusted to match the real time data.  
During data taking any deviation outside of a preset window
triggers the system to
record $4k$ samples around the trigger for later analysis. 

\section{Response to ionizing dose}

Ionizing dose tests are performed to determine the total ionizing dose
that renders the components inoperable and to determine
their annealing properties.  
These studies were performed using the
$^{60}Co$ source at the Brookhaven National Laboratory Solid State Gamma-Ray Irradiation Facility. 
DUT boards were placed at a location where the dose rate was 14~krad/h. 
Dose rates were determined
with optically stimulated luminescence (OSL) dosimeters 
provided and read by Landauer Inc. Results are 
reported in rad(Si) with an error of $\pm 6\%$ \cite{landauer}. 

Seventeen ADCs and their response to ionizing dose are presented in Table~\ref{tab:adctid}. 
All of the ADCs tested are manufactured in CMOS technology with a 180~nm feature size, except for one that
is manufactured in 350~nm feature size.  In this paper the ADC is defined as operational if the 
ADC output signal is within 5\% of the pre-irradiation amplitude, regardless of the
the current drawn. However, in most cases what we observed was a sudden failure of the
components being irradiated. As  table~\ref{tab:adctid} shows that all components withstood
doses greater than or equal to
100~kRad, with six
performing satisfactorily at doses larger than 1~$Mrad$ albeit a power increase of a factor of approximately 2. 

\begin{figure}
\centering
\includegraphics[width=0.8\columnwidth]{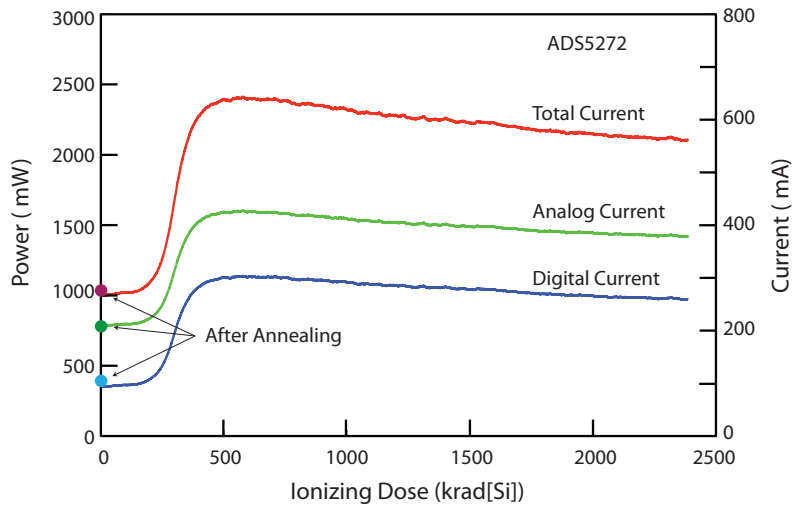}
\caption{ADS5272 current as function of ionizing dose. The device continues to digitize at ionizing doses
values keeping the signal to noise ratio. After thermal annealing at $80~^{\circ}C$ the device recovers its original 
characteristics as indicated by the solid points. }
\label{fig:annealing}
\end{figure}

\begin{figure}
\centering
\includegraphics[width=0.8\columnwidth]{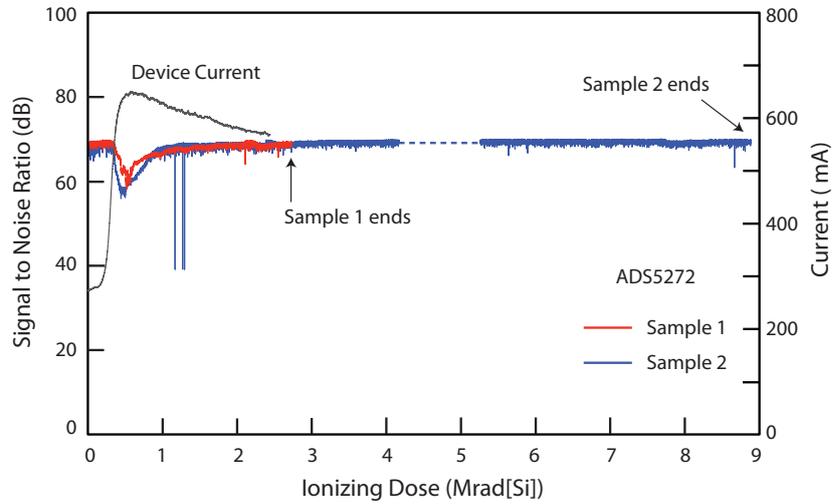}
\caption{Signal to noise ratio for two ADS5272 samples as function of ionizing dose. The increase in the device's
current is also shown. A small reduction in the signal to noise ratio is observed near the peak of current draw. Otherwise
the SNR is flat even for a region where there is an increased power consumption.  }
\label{fig:snr}
\end{figure}

Fig. \ref{fig:annealing} shows the current
increase of the ADS5272 ADC as a function of ionizing dose. The currents for both the analog and digital
sections of the ADC rise between 350 and 500 krad. 
At the higher doses
the current slowly decreases. 
The ADS5272 continues to digitize with
acceptable signal to noise ratio
as shown in Figure \ref{fig:snr}. 
However, we note that the high current consumption 
makes its use
impractical as it would require a 
power supply able to handle this large excursion.  

After irradiation, the ADC sample
was annealed at $80~^{\circ}C$ for a period of 24~h.  The ADC recovered its
original power consumption
as indicated in 
Fig. \ref{fig:annealing}.  Room temperature annealing has yielded similar results. 

Given the performance of ADS5272, we performed controlled annealing tests with twenty additional samples. 
We limited
the maximum dose to 55~krad, irradiating the components at 14~krad/h, since this
is the qualifying dose of a component for the high luminosity LHC located in
the detector outer layers. 
At this value the device current increases by 1.3~\%. The components were then
annealed at room temperature for 168~hours, followed by additional annealing at
100~$^\circ C$ for 168~hours. 
 Both the analog and digital currents returned to within 0.5\% of the
pre-irradiation values. 
This result shows that the ADS5272 has good annealing properties. This observation is
also an indication that this device is likely to be immune to enhanced low dose rate sensitivity (ELDRS). 

During ionizing tests we also determined that the ADS5272 exhibits small gain changes. 
For total integrated doses below 200~krad, the gain change is linear
with a slope of  -0.015\%/krad. Above 200~krad the ADS5272 gain changes by 3\%.
Similar measurements were performed for the ADS5294 which has
a similar response to ionizing radiation. The ADS5294 has a smaller 
gain change at -0.007\%/krad for total ionizing doses below 200~krad.

\begin{table}[hbt]
  \centering
      {\footnotesize
        \begin{tabular}{@{}ccccccccc@{}}
          \hline \hline
          ADC & Dynamic & F & Analog & Channels & $P_{total}$ per & Feature& Vendor& TID\\
           & Range & & Input Span & per  Chip& Channel & Size &  &  \\
          & [$bit$] & [$MHz$] & [$V_{p-p}$] &  & [$mW$] & (nm)& & [$kRad (Si)$] \\
          \hline
          AD9265-80 & 16 & 80 & 2 & 1 & 210 & 180 & ADI & $\sim$220 \\
          AD9268-80 & 16 & 80 & 2 & 2 & 190 & 180  & ADI & $\sim$160 \\
          AD9269-40 & 16 & 40 & 2 & 2 & 61 & 180 & ADI & $\sim$120 \\
          AD9650-65 & 16 & 65 & 2.7 & 2 & 175 & 180 & ADI & $\sim$170 \\
          AD9253-125 & 14 & 125 & 2 & 4 & 110 & 180 & ADI & $\sim$105 \\
          \hline                                                                          
          LTC2204 & 16 & 40 & 2.25 & 1 & 480 & 350 & Linear & $\sim$180 \\
          LTC2173-14 & 14 & 80 & 2 & 4 & 94 & 180 & Linear & $\sim$105 \\
          LTC2193 & 16 & 80 & 2 & 2 & 125 & 180 & Linear & $\sim$100 \\
          \hline
          ADS4245 & 14 & 125 & 2 & 2 & 140 & 180 & TI & $\sim$235 \\
          ADS6445 & 14 & 125 & 2 & 4 & 320 & 180 & TI & $\sim$210 \\
          ADS5282 & 12 & 65 & 2 & 8 & 77 & 180 & TI & $\sim$460 \\
          ADS5263 & 16 & 100 & 4 & 4 & 280 & 180  & TI & $\sim$2100 \\
          ADS5294 & 14 & 80 & 2 & 8 & 77 & 180 & TI & $\sim$1070 \\
          ADS5292 & 12 & 80 & 2 & 8 & 66 & 180 & TI & $\sim$1060 \\
          ADS5272 & 12 & 65 & 2.03 & 8 & 125 & 180 & TI & $\sim$8800 \\
          \hline  
          HMCAD1520 & 14 & 105 & 2 & 4 & 133 & 180 & Hittite & $\sim$2300 \\
          HMCAD1102 & 12 & 80 & 2 & 8 & 59 & 180& Hittite & $\sim$1730 \\
          \hline \hline 
      \end{tabular}}
  \caption{COTS ADC total ionizing dose test results. Seventeen commercial parts were tested. F is the maximum sampling frequency, and
  $P_{total}$ is the total power required to operate the component. The ADCs are from four vendors, ADI - Analog
  Devices, Linear - Linear Technologies, TI - Texas Instruments and Hittite. 
 The total ionizing dose is defines as failure to operate or with a gain change larger than 5\%  
 except for the last six ADCs that were still functional when TID test was terminated. 
    \label{tab:adctid}}

\end{table}

\section{Single Event Upset}

It is well known that high energy hadrons ($E \geq 10~MeV$) can induce single event effects in semiconductor devices. 
Unlike heavy ions, ionization in the device's critical areas is caused  by
nuclear fragments produced by the interaction of high energy hadrons with the
semiconductor nuclei.  Thus the probability of single event upset occurrence is the
product of the probabilities that a nuclear reaction occurs and that the fragments
deposit charge in a susceptible location above the SEU critical threshold. 

In an ideal measurement, SEU cross sections would be determined with hadrons with 
an energy spectrum identical to the radiation environment where the component will be used. 
This guarantees that the multiplicity and the energy of nuclear fragments 
is accurately reproduced.   Experimentally, this
is seldom possible. The best approximations to the particle spectrum presented in Fig. \ref{fig:radfield}
are neutron sources attainable at  LANSCE at Los Alamos or TSL at Uppsala \cite{lansce,tsl}. 
Particle fluxes
at these facilities are
appropriate for devices with larger cross sections ($10^{-6} - 10^{-8}~cm^{2}$)
such as RAM memories and less
suitable for devices with small cross sections
such as an ADC. For ADCs
an alternative is to
measure the SEU cross sections using high energy, high flux proton beams. 
The drawbacks are that the actual cross sections will be
different than for the expected radiation environment and a large ionizing dose is also deposited in the
devices during SEU measurements. In practice proton facilities are useful to measure processes with cross sections
of $\sim 10^{-13}~cm^{2}$  in few hours, whereas WNR-LANSCE is limited to $\sim 10^{-11}~cm^{2}$ that is achievable
in 24 hours.

\subsection{Irradiations}

\begin{table}[hbt]
  \centering
      {\footnotesize
        \begin{tabular}{@{}lcccccc@{}}
          \hline \hline
          Component & Facility & Energy &  Total Fluence & Total Dose & NIEL & Channels Read \\
                               &            & (MeV)  &  $cm^{-2}$ & krad  &  $cm^{-2}$  &     \\
           \hline
           ADS5272     & IUCF    & 205   & $5.67 \times 10^{12}$ & 338   & $5.50 \times 10^{12}$   & 1 \\
                               & IUCF    & 205   & $5.43 \times 10^{12}$ & 324   & $5.27 \times 10^{12}$   & 1 \\           
                                & IUCF    & 205   & $3.90 \times 10^{12}$ & 232   & $3.78 \times 10^{12}$   & 1 \\
           ADS5272     & MGH    & 216   & $6.75 \times 10^{12}$& 374   & $6.55 \times 10^{12}$   &  8 \\           
                               & MGH    & 216   & $4.08 \times 10^{12}$& 226   & $3.96 \times 10^{12}$   &  8 \\                  
                             & MGH    & 216   & $2.39 \times 10^{12}$& 132   & $2.31 \times 10^{12}$   &   8 \\ 
           ADS5272     & LANSCE    & $<$ 800   & $1.98 \times 10^{11}$ & $\sim$1   &   $\sim 5 \times 10^{11}$  & 1 \\                  
          \hline
           ADS5294     & MGH    & 216   & $1.55 \times 10^{12}$& 86   & $1.50 \times 10^{12}$   &  8 \\                  
                              & MGH    & 216   & $3.92 \times 10^{12}$& 217   & $3.80 \times 10^{12}$   &   8 \\    
          \hline \hline 
      \end{tabular}}
  \caption{ List of ADCs that were irradiated for single event effect measurements. For each ADC beam energy, dose, and total fluence are 
  listed. The last column shows how many channels were monitored for each irradiation. The neutron beam at LANSCE (Los Alamos) has
  an energy spectrum that mimics the atmospheric neutron spectrum. The two other facilities, the Indiana University Cyclotron Facility (IUCF) and
  Massachussetts General Hospital (MGH) deliver a mono-energetic proton beam. 
    \label{tab:seuirrad}}

\end{table}


\begin{figure}
\centering
\includegraphics[width=0.7\columnwidth]{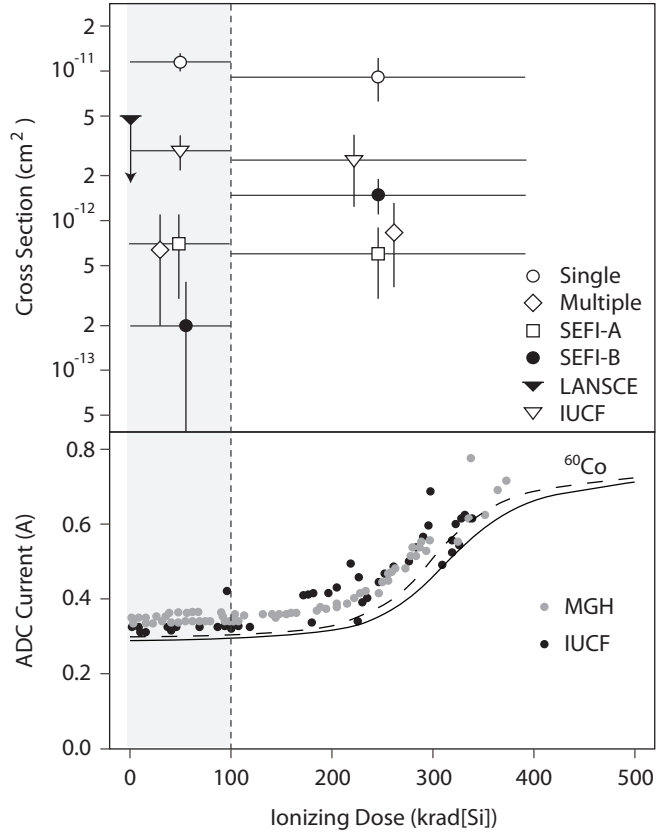}
\caption{Measured SEU cross sections for ADS5272 (top panel) and the ADC response to integrated ionizing dose (bottom panel). 
SEU cross sections for integrated dose less and larger than 100~krad are shown. SEFI A and B are defined in the text. SEU for single and multiple ADC channels are shown, together with the upper limit measured at LANSCE. The lower panel shows the increase in the ADC current as a function of integrated dose for $^{60}Co$ source and proton beam exposures.  }
\label{fig:seexsec}
\end{figure}

The two high energy proton facilities used to study single event 
effects are the Francis H. Burr Proton Therapy Center at Massachusetts General Hospital (MGH) \cite{cascio1} located in Boston, Massachusetts and the Indiana University Cyclotron Facility (IUCF) \cite{iucf} located in Bloomington, Indiana.
The MGH  is a 216~MeV proton beam facility with flux that can be tuned from
$5 \times 10^7$ to $1 \times 10^9$~$p \cdot  cm^{-2} \cdot s^{-1}$ for a circular beam spot size of approximately 2.5~cm in diameter and
a variance in flux of approximately 10\% across the spot \cite{cascio1}.  
216~MeV protons lose energy in silicon at a rate of
$dE/dx = 3.45~MeV \cdot g^{-1} \cdot cm^2$ that translates into an ionizing dose of $5.54 \times  10^{-8}$~rad(Si) per incident
proton. The IUCF is a 205~MeV proton facility with a
nominal flux of $1 \times 10^9~p \cdot cm^{-2} \cdot s^{-1}$. The beam spot is circular with a diameter of 2.5 cm.
A similar computation for ionizing dose can be done for the IUCF resulting in $5.96 \times 10^{-8}$~rad(Si) per incident proton. 
The flux at both facilities is obtained from the accelerator staff and has a 5\% uncertainty. 
Shown in the bottom panel of Fig. \ref{fig:seexsec} are the changes in the ADC current as a function of the dose
as measured at IUCF, MGH, and $^{60}Co$ facilities. 

Measurements with high energy neutrons were performed  at the WNR-LANSCE facility.  
This facility provides neutron beams
with an energy distribution that mimics the terrestrial
neutron energy spectrum. 
 The dosimetry
at LANSCE is known to 15\%. We found that during neutron irradiations, secondary charge particles produced in air or
other materials upstream of the DUT deposit  $\sim 1~krad$ over the length of the campaign. 
This was determined by placing passive
dosimeters on the components being irradiated.  

\begin{figure}
\centering
\includegraphics[width=0.7\columnwidth]{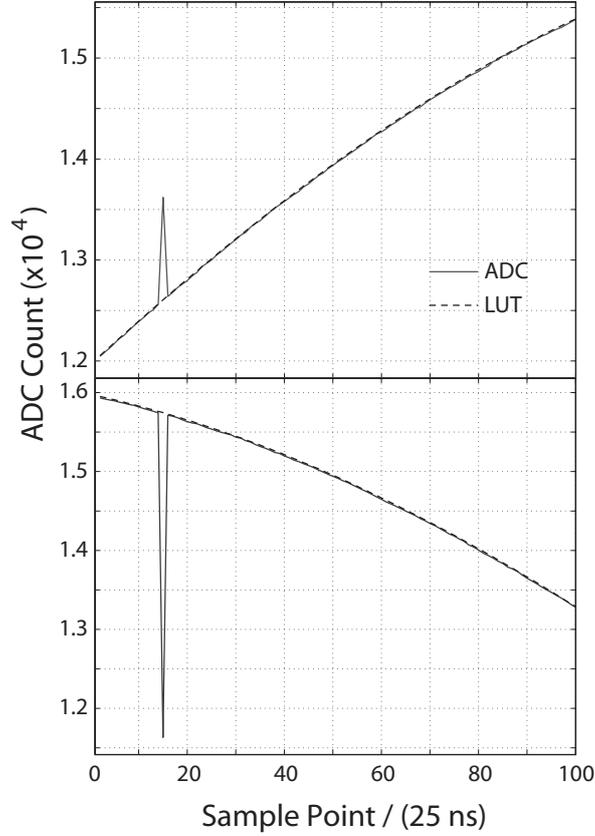}
\caption{Typical bit-flip observed during SEU tests. This type of event is observed for both ADS5272 and ADS5294. The event
is characterized by the change in the state for one clock cycle only. The top panel shows
a positive "flip" and the lower panel a negative "flip". The measured and expected values are shown by the solid and dashed lines respectively.
 }
\label{fig:glitchposneg}
\end{figure}

Proton and neutron irradiations  induce displacement damage (NIEL) due to the elastic scattering 
with $^{28}Si$ moving the $Si$ atoms from lattice sites into interstitial sites. 
The conversion of  proton or neutron fluence to 1~MeV equivalent neutron fluence can be calculated by \cite{astme}:

\begin{equation}
 \phi_{eq} = \frac{ 1}{F(1~MeV)}  \int \phi(E) F_h(E) dE
 \label{EQ1MEV}
\end{equation}

\noindent where $F_h$ is the damage function for a hadron of energy E, and $\phi(E)$ is the fluence of hadrons at 
a given energy $E$. It is 
generally accepted that $F(1~MeV)=(95 \pm 4)~mb.MeV$ \cite{astme}.  
Since all proton irradiations were performed with mono-energetic beams with $E_p \sim 200~MeV$,
only  the value for $F_h(200)$ is needed for the evaluation of the 1~MeV neutron equivalent fluence. 

Values for $F_h(200)$ can be obtained from the literature \cite{akkerman,jun,huhtinen,dale} giving an average value for
$F_h(200)/F(1~MeV)= (0.97\pm 0.05)$. For neutrons Eq. \ref{EQ1MEV} is used with  $F_h(E)$ from the ASTME tables complemented
with tables from the work of Huhtinen et al. at high energies\cite{astme, huhtinen}.  

Table \ref{tab:seuirrad} summarizes the exposures performed for the measurements of SEU cross sections. From the above considerations
both the total ionizing dose and 1~MeV neutron equivalent fluence are listed. The value calculated for LANSCE is for a
particle spectrum provided by the accelerator staff.  The provided spectrum does not include neutrons with energy below 1~MeV, and therefore
we are neglecting the contribution of lower energy neutrons. Since the damage function is very low below 0.5 MeV we don't
expect a significant contribution in the estimated 1~MeV equivalent fluence. All values
are considered below fluences where the onset of displacement damage becomes noticeable in CMOS devices.

\subsection{SEE studies}

Two categories of SEE  that influence the functionality of an ADC are data corruption  and functional 
interrupt.  Corrupted data leads to wrong information 
and single event functional interrupt (SEFI) on the other hand will lead to a longer
dead time due to the need to reset or re-initialize the ADC.  We chose
to study in detail single event effects in two candidate ADCs, the ADS5272 and the ADS5294. 
Both  ADCs have eight channels with 12 and 14 bit digitization respectively.  
The ADS5272 is the simpler device with only few user programmable registers. 
The ADS5294 is a newer generation ADC with 37 user accessible configuration registers providing the
user with a greater flexibility. 

\begin{figure}
\centering
\includegraphics[width=0.7\columnwidth]{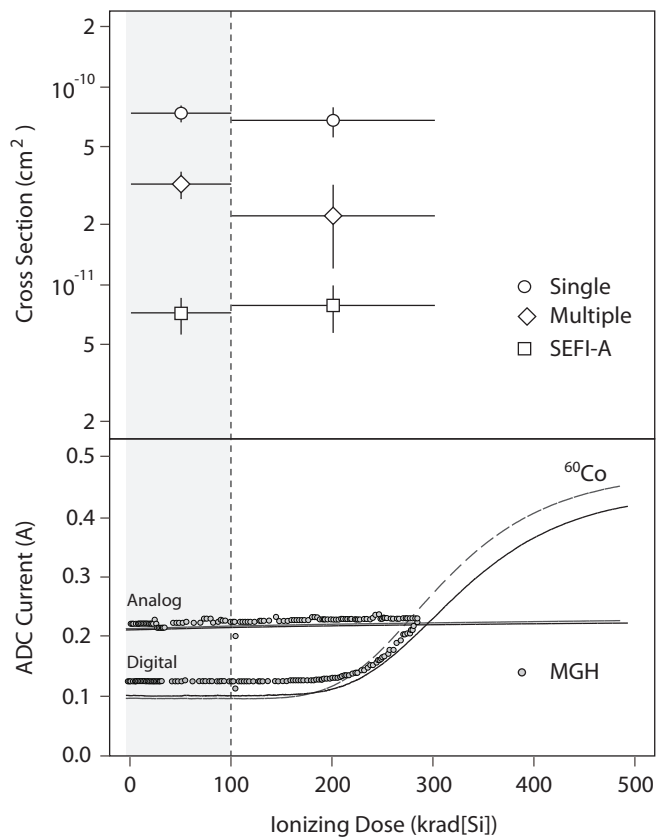}
\caption{Measured SEE cross sections for ADS5294 and the ADC response to integrated ionizing dose. 
The lower panel shows the increase in the ADC current as function of integrated dose for the analog and digital sections.
Measurements made with $^{60}Co$ source and protons are shown. SEU cross sections for integrated dose less and larger than 100~krad are shown.  SEU for single and multiple ADC channels are shown, together with SEFI cross sections.  }
\label{fig:seeads5294}
\end{figure}

For both ADCs we have identified various types of single event effects that we classify in three
categories.  The single event effect most commonly observed are transients (corrupted data)
as shown in Fig. \ref{fig:glitchposneg}. 
The effect is a change in a bit or bits
in a digital word which lasts
for
one clock cycle.  
The second type is a SEFI (defined as SEFI-A) that can be cleared by
reprogramming the user registers. The third is also a  SEFI (defined as SEFI-B) which requires a power cycle
to reinitialize the device. In analyzing the data 
we have noticed that the SEFI-B cross section is
dose dependent and hence we give the cross sections for doses below and above 100~krad of integrated
dose. 

\subsubsection{ADS5272}

The ADS5272 was tested in two configurations. Experiments conducted at LANSCE and IUCF were
performed by monitoring one ADC channel from a total of eight. The data set obtained from irradiation
campaigns at MGH was performed by reading all eight channels. In all runs, single event transient (SET)  
cross sections
were determined for events
where the difference between the actual ADC readout and expected value was larger than 31~mV or
64 ADC counts.  It should be noted that the cross section for SETs is larger than for SEFIs. In order to
increase the number of SEFIs detected
two exposures where the difference was set to 500~mV, 
(1024 ADC counts) were
performed. With less time spent recording SET events, this strategy allow us to focus mostly on functional
interrupts that happens outside of the data stream. 
In total
we have irradiated six ADCs in high energy proton beams, and one in the white
energy spectrum neutron beam. 

The top panel of figure \ref{fig:seexsec} shows the cross sections determined from the various runs. 
The largest cross section observed is for the process where one upset occurs in one of the eight ADC
channels. The measurement from the IUCF runs
is $(2.88 \pm 0.64 )$ times lower than when all eight channels are read. The exposure at LANSCE yielded
no events and therefore we only quote the
upper limit for this measurement. 
Our analysis
also shows that $\sim 5\%$ of transient events happen in more than one channel simultaneously.  
These could be caused either by simultaneous bit upsets across channels or an upset elsewhere
in the ADC control logic. 

The SEFI cross sections are lower in value.  SEFI-A is more frequent than SEFI-B for
the region of total ionizing dose lower than 100~krad. SEFI-A in this device can be cleared in 1.6~$\mu s$
by reprograming four user registers.  
Above 100~krad we observe an increase in the SEFI-B cross section by a factor of $\sim 7$
which is greater than SEFI-A cross section. 
This type of effect  has been
observed previously in different devices with notable differences in cross sections
as a function of deposited dose \cite{xapsos,schwank,seifert1,seifert2}.  Cross sections for
SEFI-A remains constant.

\subsubsection{ADS5294}

\begin{figure}
\centering
\includegraphics[width=0.7\columnwidth]{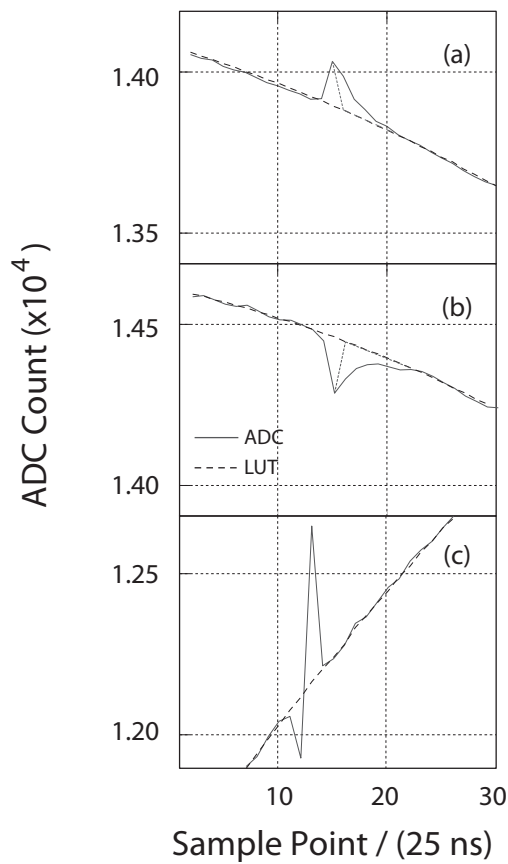}
\caption{Positive, negative and bipolar transitions observed for ADS5294.  These transitions are different than regular bit flips. 
 }
\label{fig:glitchothers}
\end{figure}

The ADS5294 is a 14 bit ADC
with 37 user configurable registers. It is a device that is significantly different than ADS5272. 
The results for ADS5294 are shown in Fig. \ref{fig:seeads5294}. The upper panel shows
the SEU cross sections measured with high energy protons at MGH, and with a detection threshold set to 
125~mV or 1024 ADC counts. 
Detailed analysis shows that  $(56 \pm 5)$\% of upsets correspond to  
a positive rising signal, 
$(39 \pm 4)$\% have a negative going signal, and we find that a small fraction of events
are bipolar glitches, $(5 \pm 1)$\%.  The analysis also shows that both positive and negative transients
have respectively
$\sim$16\% and $\sim$10\% of signals
that take 4 to 5 clock cycles to return to baseline. 
Examples of these types of transients together with bipolar transitions are
depicted in Fig. \ref{fig:glitchothers}.  The origin of this type of single event
upset is unknown. We examined the possibility that these signals are pick-ups from nearby equipment, or
the accelerator itself. We did not find any evidence that these are induced RF signal in our setup. Moreover
no signals similar to these were observed during the irradiation of ADS5272 in the same facility. 

The measured SEFI-A cross section is approximately one order of magnitude larger than
for ADS5272, which is probably due to a more complex configuration logic implemented in this ADC. 
In 70\% of SEFI-A case are instances where we observe a sudden gain change or a
constant output equal to the ADC value when the upset happened or a  null ADC output. The remaining
30\% of cases are those where the ADC output shows an oscillatory output, or a distorted 
output signal.  All SEFI-A events can be cleared in $5~\mu s$ by reprogramming the user configurable registers. 
We did not observe any dose dependence on  measured SEFI cross sections.  No associated
SEFI-B was observed in this ADC.

\section{Summary and Discussions}

This paper reports on the qualification of commercial off the self 
analog to digital converters for use in accelerator radiation environments. 
For the test and qualification, we have developed an elaborate setup that
permits us to check sample by sample if a digitized output word has been
corrupted by radiation. To achieve this, the experimental setup compares each
digitized signal with a lookup table containing the expected outcome. 
The experiment was designed having in mind the detection of single
event upsets that last one clock cycle. This detection method is limited only
by the system noise. Using this setup we thoroughly tested two
ADCs the ADS5272 and ADS5294. They were chosen from a total of 
seventeen candidates that withstood at least 100~krad of ionizing radiation. These
candidates also have desirable features
such as
effective number of bits, digitization frequency and latency. These are qualities that are
desirable for use in high energy physics experiment. 

The ADS5272 showed an exceptional recovery during annealing tests after ionizing radiation. 
Twenty samples were subjected to 55~krad total ionizing dose typically
showing a gain change of 1.3\%. After irradiation the devices were annealed
at room temperature and 100~$^\circ$C and 
returned to the pre-irradiation values to within 0.5\%. 
This makes this ADC suitable for use in applications where there are
extensive down-times, which is the case for the large hadron collider. 
It is also worth noting
that the response to ionizing radiation is the same regardless
of the radiation source. Both proton and $^{60}Co$ irradiations induce
the same response per unit dose
on both the ADS5294 and the ADS5272.  
Unlike $^{60}Co$ gamma rays, high energy protons also induce displacement
damage. We estimate that
the ADCs were exposed to
$6 \times 10^{12}~cm^{-2}$ 1~MeV neutron equivalent fluence. Although displacement
damage is not expected to cause effects
until much higher fluences in CMOS devices, the
observation that the responses are nearly identical when comparing $^{60}Co$ and proton
irradiations reassure us that there is little effect from displacement damage.

\begin{table}[hbt]
  \centering
      {\footnotesize
        \begin{tabular}{@{}lcccc@{}}
        \hline
          SEU Type & ADS5272 & ADS5272 & ADS5294 &ADS5294 \\
                  &        (<100~krad)    &  ( >100~krad) &   (<100~krad)& (>100~krad) \\
          \hline \hline
           & & & \\
          Upset in a single channel  & $(1.2 \pm 0.2) \times 10^{-11}$ & $(0.9 \pm 0.3) \times 10^{-11}$ & $(7.2 \pm 0.6) \times 10^{-11}$ & $(6.7 \pm 1.2) \times 10^{-11}$\\
          Upset in multiple channels & $(6.5 \pm 4.5) \times 10^{-13}$ & $(8.5 \pm 4.9) \times 10^{-13}$ & $(3.2 \pm 0.5) \times 10^{-11}$ & $(2.2 \pm 1.0) \times 10^{-11}$\\
          SEFI A & $(0.7 \pm 0.4) \times 10^{-12}$ &  $(0.6 \pm 0.3) \times 10^{-12}$ & $(7.1 \pm 1.5) \times 10^{-12}$  & $(7.8 \pm 2.1) \times 10^{-12}$\\
          SEFI B & $(0.2 \pm 0.2) \times 10^{-13}$ &  $(1.5 \pm 0.4) \times 10^{-12}$ &     -- & --\\
          & & & &\\
          \hline \hline 
      \end{tabular}}
  \caption{ Measured SEU cross sections for ADS5272 and ADS5294 in $cm^{-2}$ per device for total ionizing dose less and greater than 100~krad. Upset in a single channel is defined where only one of the ADC channels has an upset bit. 
SEFI A and B are defined in the text.  Quoted errors are statistical. Uncertainties in fluence will add an additional error of
5\% to the measured quantities. 
   \label{tab:seusefi}}

\end{table}

Either ADC is
well suited for use in the outer layers of high energy physics detectors where the expected
ionizing dose is low. To use these devices
the challenge is to implement mitigation strategies that will
either reduce or eliminate SEEs
to an acceptable level. For ADCs the
main problem is not SETs that produce bit flips but the single event functional interrupts, SEFIs. 
They require a more sophisticated mitigation techniques. 

To estimate the SEU rates in a real application,  we take the values of the cross sections listed
in table  \ref{tab:seusefi} and apply them to the ATLAS liquid argon calorimeter environment. 
At the position where the electronic boards are located, the expected fluence of
hadrons with $E > 20~MeV$ is $2.84 \times 10^8~cm^{-2}$ over a period of 10~hours. The system
uses 2500 ADCs to digitize the liquid argon calorimeter trigger signals.  
Under these assumptions the number
of upsets for this period for both ADCs are presented in Table \ref{tab:rates} including a safety factor of 2.  
The expectation is that for the entire calorimeter
$\sim$0.6 SEFI-B will
be observed in the 10~hours of running.  In the table we also include the typical times required to
reset the effect of the upset. 
For the LHC it is clear that any long intervention, such as a power recycle, can be made at the end of
each fill period. 
In addition, the LHC revolution includes one period of $\sim$3~$\mu s$ duration 
with no collisions every 88.924~$\mu s$. This short gap can be used to mitigate SEFIs, for example
SEFI-A in ADS5272.  
In the end, at least for this
case, the penalty paid is that in average one debilitating SEFI will happen approximately  every 20~hours of LHC operation 
at full luminosity and it
will be the experimenter's judgment if this rate is acceptable.

\begin{table}[hbt]
  \centering
      {\footnotesize
        \begin{tabular}{@{}lcccc@{}}
        \hline
          SEU Type & ADS5272 & Recovery & ADS5294 & Recovery \\
                           &                  &   time       &                 &   time  \\
          \hline \hline
           & & & & \\
          Upset in a single channel & 34 & 25~ns & 205  & 25~ns\\
          Upset in multiple channels & 2  &  25~ns & 91  & 25~ns\\
          SEFI A &   2  &  1.6~$\mu s$ &  20  & 5~$\mu s$ \\
          SEFI B  &  0.6  &  on/off & 0 & --\\
          & & & &\\
          \hline \hline 
      \end{tabular}}
  \caption{ Expected number of upsets for a scenario where $2.84 \times 10^8~cm^{-2}$ high energy hadrons with $E>20~MeV$
  strike 2500 ADCs in a period of 10~h. A safety factor of 2 was applied to the calculations. In approximately 15\% of
  cases the ADS5294 upset recovery time can be as long as 100 to 125~ns. 
   \label{tab:rates}}
\end{table}

There are 
uncertainties in the estimates of SEU rates that come from multiple sources. First, The SEU
rates are based on measurements performed with mono-energetic protons. In reality the energy spectrum of particles that compose the
environment shown in  Fig. \ref{fig:radfield}, is closer to the cosmic ray produced neutrons, similar to the LANSCE facility. 
Seifert et al. \cite{seifert}, have compared cross sections determined with mono-energetic protons and 
with the neutron energy spectrum
at LANSCE finding that 
protons consistently give a larger cross section by a factor of 1.5 to 2.0. The second source that applies to the SEFI-B in
ADS5272 is the dependence of the cross section on ionizing dose. This type of dependence has been observed previously
and could imply that for a background dominated by neutrons may have
a much lower cross section
than what is
measured with protons. The third
source of uncertainty comes from the fact that we maybe overlooking the possibility of SEU induced by the thermal
neutron capture reaction $^{10}B(n, \alpha) ^7Li$ \cite{neutron1,neutron2}. Depending on the doping concentration of 
p-type Silicon, and the flux
of thermal neutrons this reaction may induce significant numbers of
SEU. These sources of uncertainty can influence the SEU rates  and 
proper safety factors must be considered.  

We have
shown that COTS ADCs can be used in the accelerator radiation environment.  In particular, 
the ADS5272 and ADS5294 are well suited for use in the external layers of high energy physics experiments. They
are appropriate for integrated ionizing doses of
up to 200~krad, and show very good recovery during annealing periods. For the design
of electronics the main challenge is to design a mitigation strategy for single events
that is acceptable to the experimenter. We discussed the use of
pre-determined time intervals in the large hadron collider beam structure that can be used to mitigate some of functional interrupts. As new
ADCs are offered by vendors every year, a program to evaluate their susceptibility to radiation may reveal 
more ADCs candidates
that can be used in the accelerator environment.

\acknowledgments

We acknowledge the excellent service and help provided by Mr. E. Cascio at the MGH facility, Dr. B. von Przewoski  at 
IUCF and Dr. S. Wender at LANSCE. Their expert guidance was invaluable in the execution of single event upset tests. 
This work was supported in part by the Unites States Department of Energy Contract No.~DE-AC02-98CH10886.


\begin{thebibliography}{9}

\bibitem{rad1}
P. S. Winokur et al., \emph{Use of COTS microelectronics in radiation environments},  IEEE Transactions on  Nuclear Science ,vol. 46 , (1999) 1494 - 1503.

\bibitem{shupe}
M. Bosman, I. Dawson, V. Hedberg and M. Shupe
\emph{ATLAS radiation background task force summary document}, retrieved from
http://atlas.web.cern.ch/Atlas/GROUPS/PHYSICS/RADIATION/RadiationTF\_document.html

\bibitem{ibmneutron}
M. S. Gordon, P. Goldhagen, K. P. Rodbell, T. H. Zabel, H. H. K. Tang, J. M. Clem, and P. Bailey, 
\emph{Measurement of the Flux and Energy Spectrum of Cosmic-Ray Induced Neutrons on the Ground}, 
IEEE Transactions on Nuclear Science, vol. 51 (2004) 3427-3434 

\bibitem{landauer}
Landauer Inc., Online: http://www.landauer.com/

\bibitem{lansce} B. Gersey, R. Wilkins, H. Huff, R. Dwivedi, B. Takala, J. O. O'Donnell, S. A. Wender, R. C. J. Singleterry,
\emph{Correlation of neutron dosimetry using a silicon equivalent proportional counter microdosimeter and SRAM SEU cross sections for eight neutron energy spectra}. IEEE Trans. Nucl. Sci. vol. 50 (2003) 2363Ð2366.

\bibitem{tsl} A. V. Prokofiev, 
S. Pomp, J.  Blomgren, O.  Bystrom, C.  Ekstrom, O.  Jonsson, D.  Reistad,  U. Tippawan, D.  Wessman, V.  Ziemann,  M. Osterlund,
\emph{A new neutron facility for single-event effect testing}, Reliability Physics Symposium, 2005. Proceedings. 43rd Annual. 2005 IEEE International, pp. 649-695

\bibitem{cascio1} Ethan. W. Cascio, Janet. M. Sisterson, Jacob B. Flanz, and Miles. S. Wagner,
\emph{The Proton Irradiation Program at the Northeast Proton Therapy Center},
2003 IEEE Radiation Effects Data Workshop, (2003), 141 - 144

\bibitem{iucf} M.E. Rickey, M.B. Sampson, \emph{The Indiana University cyclotron facility}, 
Nuclear Instruments and Methods, Volume 97, Issue 1, 15 November 1971, Pages 65-70

\bibitem{xapsos} M. A. Xapsos, L. W. Massengill,W. J. Stapor, P Shapiro, A. B. Campbell, S. E. Kerns, K. W. Fernald and 
A. R. Knudson, \emph{Single-Event, Enhanced Single-Event and Dose-Rate Effects with Pulsed Proton Beams},
 IEEE Trans. Nuc. Sci, NS-14, 1419,(1987)

\bibitem{schwank}J. R. Schwank, M. R. Shaneyfelt,  J. A. Felix, Member, G. L. Hash, V. Ferlet-Cavrois, P. Paillet, Member, 
J. Baggio,  P. Tangyunyong, and E. Blackmore,  \emph{Issues for Single-Event Proton Testing of SRAMs}, 
IEEE Trans.  Nuc. Sci., VOL. 51, NO. 6, December 2004. 

\bibitem{seifert1} N. Seifert,   B. Gill,  S.  Jahinuzzaman, J.  Basile, V.  Ambrose,  Quan Shi, R.  Allmon and A. Bramnik, 
\emph{ Soft Error Susceptibilities of 22 nm Tri-Gate Devices}, IEEE Trans. Nuc. Sci., Vol. 59, 6, pp. 2666-2673, 2012. 

\bibitem{seifert2} S. Jahinuzzaman, B. Gill, V.  Ambrose and N. Seifert, 
\emph{ Correlating low energy neutron SER with broad beam neutron and 200 MeV proton SER for 22nm CMOS Tri-Gate devices}
IEEE IRPS pp. 3D.1.1-3D.1.6, 2013

\bibitem{astme} ASTM E722 - 09e1 Standard, \emph{Standard Practice for Characterizing 
Neutron Fluence Spectra in Terms of an Equivalent Monoenergetic Neutron Fluence for Radiation-Hardness 
Testing of Electronics}, October, 2009. 

\bibitem{akkerman} A. Akkerman, J. Barak, M.B. Chadwick, J. Levinson, M. Murata, and Y. Lifshitz,
\emph{Updated NIEL calculations for estimating the damage induced by particles and $\gamma$-rays in Si and GaAs}
Radiation Physics and Chemistry 62 (2001) 301-310

\bibitem{jun} Insoo Jun, Michael A. Xapsos, Scott R. Messenger, Edward A. Burke, Robert J. Walters, Geoff P. Summers, and
Thomas Jordan,
\emph{Proton Nonionizing Energy Loss (NIEL) for Device Applications}
IEEE Transactions on Nuclear Science, Vol. 50, No. 6, (2003) 1924

\bibitem{huhtinen} M. Huhtinen and P.A. Aarnio, 
\emph{ Pion induced displacement damage in silicon devices} 
Nucl. Instr. Meth. In Phys. Res. A, vol 335, pp 581-582, 1993. 

\bibitem{dale} G. R. Hopkinson,  C. J. Dale, and P. W. Marshall,
\emph{Proton Effects in Charge-Coupled Devices}
IEEE Transactions on Nuclear Science, vol. 43, No. 2, (1996) 614

\bibitem{seifert} N. Seifert, S. Jahinussaman, J. Basile, Q. Shi, R. Allmon and A. Braknik, \emph{Soft Error Susceptibilities of 22 nm Tri-Gate
Devices}, IEEE Trans. on Nucl. Science, vol 99, No. 6 (2012) 2666- 2673

\bibitem{neutron1}A Vazquez-Luque, J. Marin, J.A. Terron,, M.  Pombar, R.  Bedogni, F.  Sanchez-Doblado and F. Gomez. 
\emph{Neutron Induced Single Event Upset Dependence on Bias Voltage for CMOS SRAM With BPSG},  IEEE Trans. on Nuclear Physics, vol. 60, (2013) 4692 - 4696

\bibitem{neutron2}Yutaka Arita, Mikio Takai, Izumi Ogawa and Tadafumi Kishimoto, \emph{Experimental Investigation of Thermal Neutron-Induced Single Event Upset in Static Random Access Memories},  Japanese. J. Appl. Phys. 40 (2001) L151 doi:10.1143/JJAP.40.L151. 





\end{thebibliography}
\end{document}